\documentclass[a4paper,twoside,10pt]{article}
\usepackage{graphicx,publaob}

\def\papertype{~}
\begin{document}

% Title
\title{COMPARISON OF GROUND-BASED AND GAIA PHOTOMETRY OF ASTROMETRIC RADIO SOURCES}

% Authors
\authors{Z. MALKIN$^{1,2}$}

% Addresses and e-mails
\address{$^1$Pulkovo Observatory, Pulkovskoe shosse 65, St. Petersburg 196140, Russia}
\Email{malkin}{gaoran}{ru}
\address{$^2$Kazan Federal University, Kremlyovskaya street 18, Kazan 420008, Russia}

% Running titles
\markboth{COMPARISON OF GROUND-BASED AND GAIA PHOTOMETRY}{Z. MALKIN}

% Abstract
\abstract{
A comparison was made between $Gaia$ magnitudes and magnitudes obtained
from ground-based observations for astrometric radio sources .
The comparison showed that these magnitudes often not agree well.
There may be several reasons for this disagreement.
Nevertheless, such an analysis can serve as an additional filter for verification of the object cross-identification.
On the other hand, it can help to detect possible errors in optical magnitudes of astrometric radio sources coming
from unreliable or inconsistent data sources.
}

%%%%%%%%%%%%%%%%%%%%%%%%%%%%%%%%%%%%%%%%%%%%%%%%%%%%%%%%%%%%%%%%%%%%%%%%%%%%%%%%%%%%%%

\section{INTRODUCTION}

Comparison of the VLBI and $Gaia$ positions of astrometric radio sources gives us an opportunity to investigate
both systematic errors of both catalogs and the physical structure and evolution of objects observed by optical
and radio techniques.
However, the reliability of the results of such studies depends on the reliability of cross-identification between
objects in VLBI and $Gaia$ catalogs.
Although for most sources the cross-identification looks straightforward, this is is not always the case.
Errors in cross-identification introduce additional noise and outliers in the position difference statistics,
which may impact the result and conclusions of these studies.

Different authors use different strategies for cross-identification between the objects in radio and optical catalogs
(Mignard et al. 2016, Petrov and Kovalev 2017, Lindegren et al. 2018, Malkin 2018, Makarov et al. 2019, Liu et al. 2020).
They all are based on analysis of the differences between VLBI and $Gaia$ source position, sometimes along with
other criteria, such as quality of the $Gaia$ astrometric solution (Mignard et al. 2016, Lindegren et al. 2018)
or object density in the vicinity of the source (Petrov and Kovalev 2017).
Some of these supplement criteria may be disputable, but a comparison and discussion of cross-identification
methods described in the literature is beyond the scope of this paper.

The present study is aimed at an investigation of a possibility of using $Gaia$ and ground-based photometry
as a supplement criterium to verify the cross-identification based on an analysis of the difference between
radio and optics positions.

%%%%%%%%%%%%%%%%%%%%%%%%%%%%%%%%%%%%%%%%%%%%%%%%%%%%%%%%%%%%%%%%%%%%%%%%%%%%%%%%%%%%%%%%%%%%%%%%%%%%%%%%%%%%%%

\section{COMPARISON OF GROUND-BASED AND GAIA PHOTOMETRY}
\label{sect:comparison}

In this study, the $G$ magnitudes given in the $Gaia$ DR2 release were compared to ground-based magnitudes
provided in the OCARS catalog\footnote{http://www.gaoran.ru/english/as/ac\_vlbi/ocars\_m.txt} (Malkin 2018).
There are 8197 sources in common between the $Gaia$ DR2 and OCARS.
7179 of them have photometry data in OCARS in at least one of four bands: $V$, $r$, $R$, and $i$,
close to $Gaia$ $G$ band.
The latter sources were used in this work.

\begin{figure}
\centering
\includegraphics[width=\textwidth]{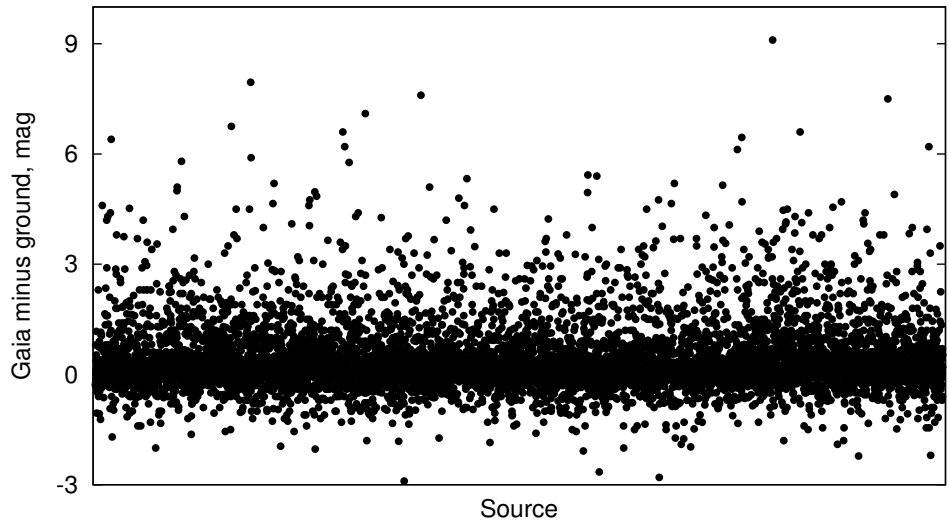}
\caption{Differences between $Gaia$ $G$ and ground-based magnitudes for 7179 common sources between the $Gaia$ DR2 and OCARS catalogs.}
\label{fig:gaia-ocars}
\end{figure}

OCARS magnitudes are taken from different data sources:
Sloan Digital Sky Survey\footnote{http://www.sdss.org} (SDSS),
NASA/IPAC Extragalactic Database\footnote{http://ned.ipac.caltech.edu} (NED),
SIMBAD\footnote{http://simbad.u-strasbg.fr/simbad/} database managed by the Centre de Donnees astronomiques de Strasbourg (CDS),
the Million Quasars (Milliquas) catalog\footnote{https://heasarc.gsfc.nasa.gov/W3Browse/all/milliquas.html},
and Large Quasar Astrometric Catalogue\footnote{https://cdsarc.unistra.fr/viz-bin/cat/J/A\%2bA/624/A145} (LQAC),
in the order of preference.
The contribution of two latter catalogs is small.
The data in these catalogs and data bases are not always consistent and reliable.
To mitigate this impact, the average magnitude of $V, r$, $R$, and $i$ was  used in this work.
The magnitude differences $Gaia$ $G$ minus OCARS for all 7179 common sources are shown in Fig.~\ref{fig:gaia-ocars}.
One can see that they can be both positive and negative in the interval from about $-3^m$ to about $9^m$.

\begin{figure}
\centering
\includegraphics[width=\textwidth]{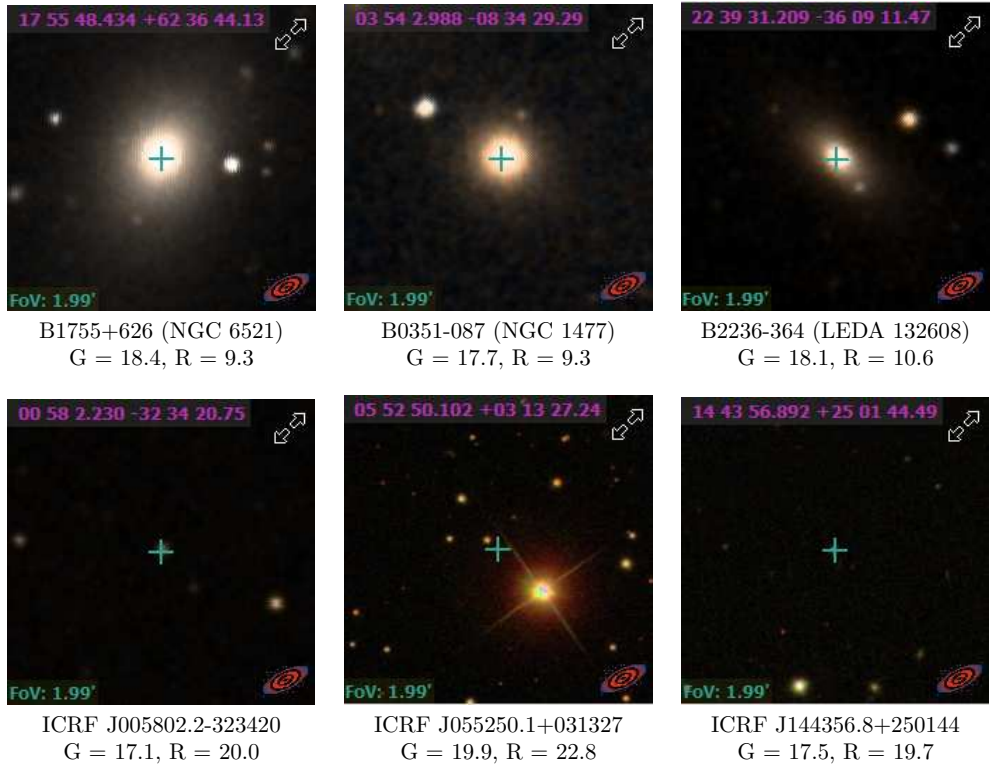}
\caption{Top row: three objects with large negative $Gaia$ minus OCARS optical magnitude difference. They all are large galaxies.
  Bottom row: three faint objects with large positive $G$--$R$ difference.
  The images are provided by the Aladin Sky Atlas (\mbox{https://aladin.u-strasbg.fr/}).}
\label{fig:images}
\end{figure}

The positive magnitude differences can generally be explained by the fact that corresponding objects are extended galaxies
hosting AGN observed by VLBI.
Their ground-based (OCARS) magnitudes are related to the integrated flux from the entire galaxy
or its wide central region, whereas $Gaia$ most probably observes the brightest point-like center of the galaxy.
Three typical objects of this type are shown in Fig.~\ref{fig:images}.
Therefore, positive magnitude difference is mostly a trivial case.

\begin{figure}
\centering
\includegraphics[width=\textwidth]{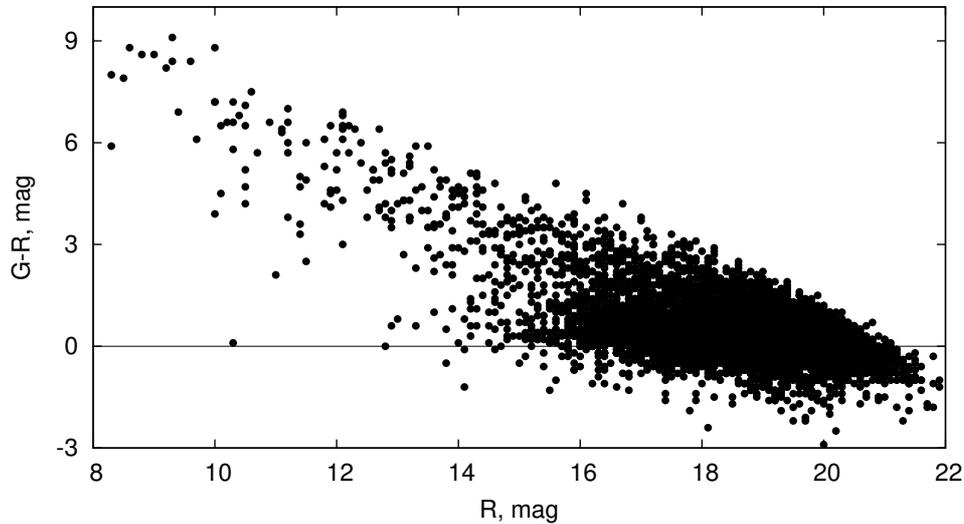}
\caption{Differences between $Gaia$ $G$ and ground-based $R$ magnitudes as function of $R$.}
\label{fig:gaia-ocars_rg}
\end{figure}

To check this conclusion, the same data were re-organized and depicted in Fig~\ref{fig:gaia-ocars_rg}, where the magnitude difference $G$--$R$
is shown as a function of $R$.
One can see that optically faint objects show relatively small magnitude difference $Gaia$ minus OCARS, while bright objects show much large
differences.
Most probably, this is a consequence of the $Gaia$ observing mode for which the faint galaxies are observed as a single object,
unlike large (extended) galaxies.
Detailed discussion of observation of galaxies with $Gaia$ is given by de Souza et al. 2014. 

More interesting is a case of negative magnitude difference, which is the subject of a special investigation
for each source that shows such a feature.
There may be various reasons for this:
\begin{itemize}
\item The observed magnitude difference may be a result of wrong cross-identification of the astrometric
radio source with $Gaia$ object.
\item The observed magnitude difference may be a result of a source optical variability.
The photometry data in $Gaia$ and OCARS were generally made at different epochs as discussed
by Damljanovi\'c et al. 2017.
Results of optical monitoring of ICRF sources made by, e.g., Taris et al. 2016, showed the spread of
the brightness variations may exceeds two magnitudes.
\item The observed magnitude difference may be a result of wrong cross-identification of the radio source
with an optical object in one of the catalogs contributed to OCARS.
In the course of the present work about a half of dozen of such cases was found.
\end{itemize}

Of course, the same considerations can be applied to positive $Gaia$ minus OCARS magnitude differences.

%%%%%%%%%%%%%%%%%%%%%%%%%%%%%%%%%%%%%%%%%%%%%%%%%%%%%%%%%%%%%%%%%%%%%%%%%%%%%%%%%%%%%%%%%%%%%%%%%%%%%%%%%%%%%%

\section{CONCLUSION}
\label{sect:conclusion}

The goal of this work was to investigate whether a comparison of the optical magnitudes of astrometric radio
sources with the magnitudes obtained from ground-based (or other wide-field) observations can be useful to improve
the reliability of cross-identification of $Gaia$ objects with radio sources included in VLBI astrometric catalogs.
It was found that the direct comparison of $Gaia$ and ground-based magnitudes can hardly be practical for this purpose.
A typical case is a bright galaxy for which VLBI and $Gaia$ positions are measured for the optically and radio
strong central point-like area (AGN), whereas the ground-based magnitudes are related to the integral galaxy flux.

Therefore, comparison of the $Gaia$ and OCARS magnitudes cannot be used straightforward for check of object
cross-identification in $Gaia$ and VLBI catalogs.
However, in some cases, this may be useful to detect a wrong cross-identification, especially in case of negative
$Gaia$ minus OCARS magnitude differences.
Besides, this comparison allows us to identify possible inconsistencies or even errors in a ground-based photometry
data taken from different, not always reliable, catalogs and data bases.
In particular, in the course of this work, several OCARS sources with erroneous magnitudes were detected and
corrected or flagged in OCARS.

In any case, a further deeper investigations in this direction may be useful to improve the reliability of the
object cross-identification in VLBI and $Gaia$ catalogs.

%%%%%%%%%%%%%%%%%%%%%%%%%%%%%%%%%%%%%%%%%%%%%%%%%%%%%%%%%%%%%%%%%%%%%%%%%%%%%%%%%%%%%%%%%%%%%%%%%%%%%%%%%%%%%%

%\section*{Acknowledgment}

\vskip 6.5mm
\centerline{\titlefont Acknowledgments}
\vskip 3mm

This work has made use of data from the European Space Agency (ESA) mission $Gaia$\footnote{https://www.cosmos.esa.int/gaia},
processed by the $Gaia$ Data Processing and Analysis
Consortium\footnote{https://www.cosmos.esa.int/web/gaia/dpac/consortium} (DPAC).
Funding for the DPAC has been provided by national institutions, in particular the institutions participating
in the $Gaia$ Multilateral Agreement.

This research has made use of the SIMBAD database and Aladin Sky Atlas operated at at the Centre de Donn\'ees
Astronomiques\footnote{https://cds.u-strasbg.fr/} (CDS), Strasbourg, France,
and the SAO/NASA Astrophysics Data System\footnote{https://ui.adsabs.harvard.edu/} (ADS).

This research has made use of the SAO/NASA Astrophysics Data System\footnote{https://ui.adsabs.harvard.edu/} (ADS).

This work was partly supported by the Russian Government program of Competitive Growth of Kazan Federal University.

The author is grateful to Sergey Klioner for useful discussion and to the anonymous referee for suggestions
on improving the manuscript.

%%%%%%%%%%%%%%%%%%%%%%%%%%%%%%%%%%%%%%%%%%%%%%%%%%%%%%%%%%%%%%%%%%%%%%%%%%%%%%%%%%%%%%%%%%%%%%%%%%%%%%%%%%%%%%

\references

de Souza, R. E., Krone-Martins, A., dos Anjos, S., Ducourant, C., Teixeira, R. : 2014, \journal{Astron. Astrophys.}, \vol{568}, A124.
 
Damljanovi\'c, G, Taris, F., Andrei, A. : 2017, \journal{Astrometry and Astrophysics in the Gaia sky, Proc. IAU}, \vol{330}, 88.
 
Lindegren, L., Hern\'andez, J., Bombrun, A., et al. : 2018, \journal{Astron. Astrophys.}, \vol{616}, A2.

Liu, N., Lambert, S. B., Zhu, Z., Liu, J. C. : 2020, \journal{Astron. Astrophys.}, \vol{634}, A28.

Makarov, V. V., Berghea, C. T., Frouard, J., Fey, A., Schmitt, H. R. : 2019, \journal{Astrophys. J.}, \vol{873}, 132.

Malkin, Z. : 2018, \journal{Astrophys. J. Suppl. Series}, \vol{239}, 20.

Mignard, F., Klioner, S., Lindegren, L., et al. : 2016, \journal{Astron. Astrophys.}, \vol{595}, A5.

Petrov, L., Kovalev, Y. Y. : 2017, \journal{Mon. Not. R. Astron. Soc.}, \vol{467}, L71.

Taris, F., Andrei, A., Roland, J., Klotz, A., Vachier, F., Souchay, J. : 2016, \journal{Astron. Astrophys.}, \vol{587}, A112.

\endreferences

\end{document}